\documentclass[conference]{IEEEtran}
\IEEEoverridecommandlockouts
\usepackage{cite}
\usepackage{amsmath,amssymb,amsfonts}
\usepackage{algorithmic}
\usepackage{graphicx}
\usepackage{textcomp}
\usepackage{xcolor}
\usepackage{orcidlink}
\usepackage{listings}
\definecolor{codegreen}{rgb}{0,0.6,0}
\definecolor{codegray}{rgb}{0.5,0.5,0.5}
\definecolor{codepurple}{rgb}{0.58,0,0.82}
\definecolor{backcolour}{rgb}{0.95,0.95,0.92}

\lstdefinestyle{mystyle}{
    backgroundcolor=\color{white},   
    commentstyle=\color{codegreen},
    keywordstyle=\color{magenta},
    numberstyle=\tiny\color{codegray},
    stringstyle=\color{codepurple},
    basicstyle=\ttfamily\footnotesize,
    breakatwhitespace=false,         
    breaklines=true,                 
    captionpos=b,                    
    keepspaces=true,                 
    numbersep=5pt,                  
    showspaces=false,                
    showstringspaces=false,
    showtabs=false,                  
    tabsize=2,
    frame=single, 
    postbreak=\mbox{\textcolor{red}{$\hookrightarrow$}\space}
}
\definecolor{eclipseStrings}{RGB}{42,0.0,255}
\definecolor{eclipseKeywords}{RGB}{127,0,85}
\colorlet{numb}{magenta!60!black}

\lstdefinelanguage{json}{
    basicstyle=\normalfont\ttfamily,
    commentstyle=\color{eclipseStrings}, 
    stringstyle=\color{eclipseKeywords}, 
    numberstyle=\scriptsize,
    stepnumber=1,
    numbersep=8pt,
    showstringspaces=false,
    frame=single,
    backgroundcolor=\color{white}, 
    string=[s]{"}{"},
    comment=[l]{:\ "},
    morecomment=[l]{:"},
    literate=
        *{0}{{{\color{numb}0}}}{1}
         {1}{{{\color{numb}1}}}{1}
         {2}{{{\color{numb}2}}}{1}
         {3}{{{\color{numb}3}}}{1}
         {4}{{{\color{numb}4}}}{1}
         {5}{{{\color{numb}5}}}{1}
         {6}{{{\color{numb}6}}}{1}
         {7}{{{\color{numb}7}}}{1}
         {8}{{{\color{numb}8}}}{1}
         {9}{{{\color{numb}9}}}{1}
}

\definecolor{bluekeywords}{rgb}{0.13,0.13,1}
\definecolor{greencomments}{rgb}{0,0.5,0}
\definecolor{turqusnumbers}{rgb}{0.17,0.57,0.69}
\definecolor{redstrings}{rgb}{0.5,0,0}

\lstdefinelanguage{pseudo}
    {morekeywords={new, type, of, AND, OR, for, in, do, begin, end, fun, if, then, else, while, each},
    keywordstyle=\color{bluekeywords},
    sensitive=false,
    morecomment=[l][\color{greencomments}]{///},
    morecomment=[l][\color{greencomments}]{//},
    morecomment=[s][\color{greencomments}]{{(*}{*)}}
    }
    
\lstdefinelanguage{vtemp}
    {morekeywords={\#if, \#else, \#end},
    keywordstyle=\color{bluekeywords},
    sensitive=false,
    alsoletter=\#,
    morecomment=[l][\color{greencomments}]{///},
    morecomment=[s][\color{redstrings}]{{\$\{}{\}}}
    }
\usepackage{tikz}
\newcommand*\circled[1]{\tikz[baseline=(char.base)]{
            \node[shape=circle,draw,inner sep=2pt] (char) {#1};}}

\def\BibTeX{{\rm B\kern-.05em{\sc i\kern-.025em b}\kern-.08em
    T\kern-.1667em\lower.7ex\hbox{E}\kern-.125emX}}

\makeatletter
\newcommand{\linebreakand}{%
  \end{@IEEEauthorhalign}
  \hfill\mbox{}\par
  \mbox{}\hfill\begin{@IEEEauthorhalign}
}
\makeatother
\begin{document}

\title{Code Generation for Machine Learning using Model-Driven Engineering and SysML\thanks{This work has been partially supported and funded by the Austrian Research Promotion Agency (FFG) via the" Austrian Competence Center for Digital Production" (CDP) no. 881843.}
}

\author{\IEEEauthorblockN{1\textsuperscript{st} Simon Raedler~\orcidlink{0000-0003-1491-7170}}
\IEEEauthorblockA{\textit{Department of Computer Science} \\
\textit{TUM School of Computation, Information and Technology}\\
Technical University of Munich, Germany \\
simon.raedler@tum.de}
\and
\IEEEauthorblockN{2\textsuperscript{nd} Matthias Rupp}
\IEEEauthorblockA{\textit{Department of Computer Science} \\
University of Applied Sciences Vorarlberg\\
Dornbirn, Austria}
\linebreakand
\IEEEauthorblockN{3\textsuperscript{rd} Eugen Rigger}
\IEEEauthorblockA{\textit{Department of Digital Engineering} \\
Zumtobel Lighting GmbH\\
Dornbirn, Austria}
\and
\IEEEauthorblockN{4\textsuperscript{th} Stefanie Rinderle-Ma~\orcidlink{0000-0001-5656-6108}}
\IEEEauthorblockA{\textit{Department of Computer Science} \\
\textit{TUM School of Computation, Information and Technology}\\
Technical University of Munich, Germany \\
stefanie.rinderle-ma@tum.de}
}

\maketitle

\begin{abstract}
Data-driven engineering refers to systematic data collection and processing using machine learning to improve engineering systems.
Currently, the implementation of data-driven engineering relies on fundamental data science and software engineering skills.
At the same time, model-based engineering is gaining relevance for the engineering of complex systems.
In previous work, a model-based engineering approach integrating the formalization of machine learning tasks using the general-purpose modeling language SysML is presented.
However, formalized machine learning tasks still require the implementation in a specialized programming languages like Python.
Therefore, this work aims to facilitate the implementation of data-driven engineering in practice by extending the previous work of formalizing machine learning tasks by integrating model transformation to generate executable code.
The method focuses on the modifiability and maintainability of the model transformation so that extensions and changes to the code generation can be integrated without requiring modifications to the code generator.
The presented method is evaluated for feasibility in a case study to predict weather forecasts.
Based thereon, quality attributes of model transformations are assessed and discussed.
Results demonstrate the flexibility and the simplicity of the method reducing efforts for implementation. 
Further, the work builds a theoretical basis for standardizing data-driven engineering implementation in practice.
\end{abstract}

\begin{IEEEkeywords}
Model-Driven Engineering, Machine Learning, Model Transformation, SysML
\end{IEEEkeywords}

\section{Introduction}
The attractiveness of machine learning and data mining in engineering has been increasing for years, as seen in the number of publications on machine learning and data mining\cite{dogan_machine_2021}.
In technical product development, the application of machine learning for making informed decisions is called data-driven engineering\cite{trauer_data-driven_2020}.
The complexity of technical product development is increasing due to the number of components, functions, and interactions of systems. 
This in turn leads to an increasing need for Model-Based (Systems) Engineering (MBE) techniques, which are promising to manage the complexity due to different system modeling methods proven in practice\cite{beihoff_world_2014,huldt_state--practice_2019,madni_model-based_2018}.
However, MBE techniques focus on formalizing knowledge rather than processing data to generate valuable insights.
Therefore, efforts are required to integrate data-driven engineering into technical product development and support to make informed decisions.
Consequently, the formalized integration of data-driven engineering or machine learning into MBE approaches is necessary. 
The authors previous work introduced a model-based formalization of machine learning tasks based on the systems modeling language SysML\cite{radler_integration_2022}.
Although this approach supports the formalization of machine learning tasks using SysML, a gap exists between the formalized knowledge within the SysML model and the actual implementation in dedicated programming languages such as Python.
In this respect, this work aims to introduce a method for the automatic generation of machine learning code to reduce the duplication of effort for formalizing and implementing machine learning tasks and to extend the model-based approach to a model-driven approach.
Consequently, the following research questions are elaborated in this work: 

\begin{enumerate}
    \item[RQ1] Which model properties can be used in the context of model-driven engineering to automatically derive a machine learning model?
    \item[RQ2] What means of software engineering allows to extend and maintain the machine learning code derivation without changes in the model transformation?
\end{enumerate}

As a result, this work presents a method that facilitates the implementation of machine learning code by deriving the formalization of machine learning tasks in SysML using a mapping mechanism that completes code snippets with properties and contexts from SysML and introduced stereotypes.
From a more general point of view, this work contributes by improving the efficiency and effectiveness~\cite{clarkson_design_2005} of the development of machine learning in the context of systems engineering. 
The method is implemented and assessed for feasibility based on a use case involving an online dataset for weather forecasting including sensor data.
The source of the implementation and evaluation is available online.
Furthermore, an evaluation and justification with regard to the quality characteristics of the model transformation\cite{van_amstel_assessing_2012} is conducted.
The reminder of the paper is as follows.
First, relevant background on MBE and the previously introduced approach to model machine learning concerns using SysML is depicted.
Second, a method is introduced, allowing to derive machine learning code using model transformation techniques.
Next, an evaluation based on an open dataset is presented.
Finally, the results are discussed, future work is highlighted, and a conclusion is presented.

\section{Background}
\label{sec:background}
In the following, relevant background concerning Model-Based Engineering (MBE), SysML, and a basic understanding of the machine learning modeling method published in \cite{radler_integration_2022} are presented. Additionally, related work is discussed.

\subsection{Model-Based Engineering \& Model Transformation}
The core of MBE includes the pillar concepts of models, metamodels, and model transformation\cite{brambilla_model-driven_2017}.
Depending on the application domain, the involved engineering concepts, e.g., software or hardware, and the degree of automation, various acronyms are typically used for the concepts of MBE\footnote{See \url{https://modeling-languages.com/clarifying-concepts-mbe-vs-mde-vs-mdd-vs-mda/} for a discussion.}.
Model transformation can be characterized as the mapping between one or multiple input and output models.
The mapping itself is defined on metamodels and not on the actual instances of a metamodel (model) to allow for reuse and generality. 
Model transformation aims to achieve the highest degree of automation by mapping artifacts\cite{brambilla_model-driven_2017}.
The transformation can either be programmed manually using any programming language or using appropriate languages provided by the model-driven software engineering domain, e.g., ATL\footnote{\label{footnote}\url{https://www.eclipse.org/atl/}}, Epsilon\footref{footnote}, etc.
Model transformations can be classified as model-to-model or model-to-text transformations, depending on whether the transformation output is a model or text/code~\cite{brambilla_model-driven_2017}.

\subsection{Machine Learning Task Formalization using SysML}\label{sub:MLFormalization}
\begin{figure}
    \centering
    \includegraphics[width=\columnwidth]{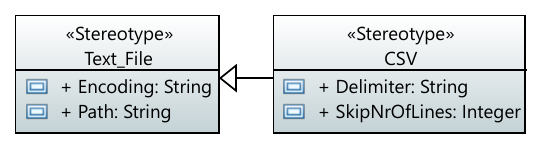}
    \caption{Sample stereotype defining a text and CSV data source.}
    \label{fig:CSVStereotype}
\end{figure}

SysML is a general-purpose modeling language allowing to describe a system of interest with machine-readable artifacts.
In previous work, we introduced the concept of machine learning task definition based on an extension of the SysML metamodel using stereotypes and the definition of semantics to interpret the model~\cite{radler_integration_2022}.
A stereotype is a concept that allows the semantics of a metamodel to be extended to include specific properties suitable for a particular purpose and to which any block applying the stereotype must conform.
A block in SysML represents a specific system, abstraction of a part of a system, or components, among others.
The particular purpose of a block in the preliminary work is the representation of a specific task or subtask of a machine learning definition, e.g., data preprocessing or more specific, datetime conversion.
In this respect, each stereotype abstracts a specific function or set of functions applied to a specific input value, e.g., the loading of a CSV file is abstracted using a stereotype defined in Figure \ref{fig:CSVStereotype}.
The properties of a stereotype are the mandatory parameters of the abstracted function and have to be defined.
Additionally, properties of a stereotype can be inherited, allowing to define specific attributes only once, e.g., \textit{Path} attribute is valid for a text and CSV file.

\begin{figure}
    \centering
    \includegraphics[width=\columnwidth]{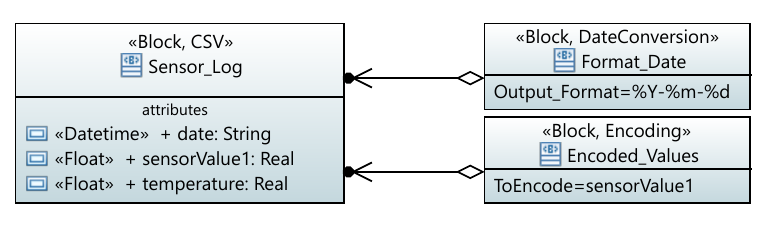}
    \caption{Sample application of stereotypes and semantic integration.}
    \label{fig:SampleApplicationML}
\end{figure}

Figure \ref{fig:SampleApplicationML} depicts the application of the defined stereotypes.
The block association indicates that a specific block that is \textit{part of} another block can be interpreted as input.
Date conversion, for example, is applied to the \textit{Sensor\_Log} data source of type \textit{CSV} with a defined \textit{Output\_Format}.
Since only the \textit{date} attribute is suitable for the date conversion, no further details are required.
Still, if further details are necessary, the modeling can be extended by adding additional attributes, e.g., selecting the correct input value, etc.

The modeling of the specific machine learning tasks is based on block definition diagrams, which are a means of structural modeling.
To specify an execution order for a set of functions abstracted behind a block, behavior diagrams, especially state diagrams, are used.
More precisely, each state of a state diagram is connected to a previously-defined machine learning block.
The connection is established using a custom stereotype.
With the connection of blocks to the state diagram, the execution order of the method is defined:
For this, each task is specified with a sequential execution order, allowing implementation or deriving machine learning code.

\subsection{Related Work and Research Gaps}
The concept of model-driven software engineering with a special focus on machine learning concerns can be found in literature\cite{moin_model-driven_2022,bhattacharjee_stratum_2019,kusmenko_engineering_2019}. 

In~\cite{moin_model-driven_2022}, an extension of the CPS modeling framework ThingML~\cite{harrand_thingml_2016} called ThingML+ is proposed. 
The extension ThingML+ allows to model machine learning artifacts using a textual domain-specific language.
The extension focuses on modeling supervised machine learning, and the Xtext-based transformation generates both Java and Python code.
Instead of a standardized general-purpose modeling language like SysML, a custom domain-specific modeling language is used. 
Customization and extension of the code generation or adding additional machine learning algorithms require extending the source code by adding specific algorithms.

In~\cite{bhattacharjee_stratum_2019}, a platform supporting the integration of machine learning in a cloud application by experts called ``Stratum'' is proposed.
The domain-specific modeling language allows the modeling of machine learning pipelines and models. 
The models and functions can be enriched with parameters, such as hyper-parameters for the learning method.
Various machine learning frameworks are integrated and code can be generated.
A graphical modeling interface is available by using WebGME\footnote{\url{https://webgme.org/}} as a base.
The extension and customization of the code generation require code extensions.
A shortcoming of the method is the stiffness of the code snippets for the code generation, making it hard to use the generator for approaches other than the proposed case study.

In~\cite{kusmenko_engineering_2019}, textual modeling is used to describe neural networks.
The approach mainly focuses on artificial neural networks.
We have observed that a considerable amount of work is required to add more functions.
Additionally, the integration of various data sources is limited.

Apart from related scientific work, frameworks such as KNIME\footnote{\url{https://www.knime.com/}} or RapidMiner\footnote{\url{https://rapidminer.com/}} can be considered as related approaches.
However, these approaches solely rely on machine learning concerns while their embedding into product development, such as MBE approaches, is not provided.
Furthermore, the higher degree of freedom to formalize and document specific machine learning concerns requires less rigidity in terms of extensibility and adaptability.

Summarizing the analysis of existing literature, machine learning code generation based on model-driven methods is under development and state of the art\cite{burgueno_mde_2021}.
The actual approaches mainly rely on custom domain-specific languages that define machine learning tasks using models.
However, the given approaches are stiff regarding extensions due to the encapsulation of the machine learning algorithms in the source code of the code generation.
Additionally, the integration of knowledge from intersecting domains is not given, making it hard to synchronize changes or to transfer knowledge.
Last but not least, the approaches often propose new modelling languages, which in turn might limit users to the already scarce Data Scientists\cite{radler_survey_2022}, instead of extending the modelling languages used in other fields.
Accordingly, using a modeling language that is already known may make sense.


\section{Method}
\label{sec:method}
In preliminary work, a method to describe all relevant information for implementing a machine learning approach using SysML is defined\cite{radler_integration_2022}. Particularly, the model represents all information concerning the composition of various relevant systems, their related data collection and the formalization of relevant data transformation and machine learning-related tasks on a single step (subtask) level. Additionally, the execution order of the machine learning tasks in the implementation is formalized using state diagrams. Each state of the diagram describes a set of sub-activities, e.g., a sequence of python functions with a dedicated purpose, such as the transformation of \textit{Datetime} into another format. 

\begin{figure*}[ht]
    \centering
    \includegraphics[width=\textwidth]{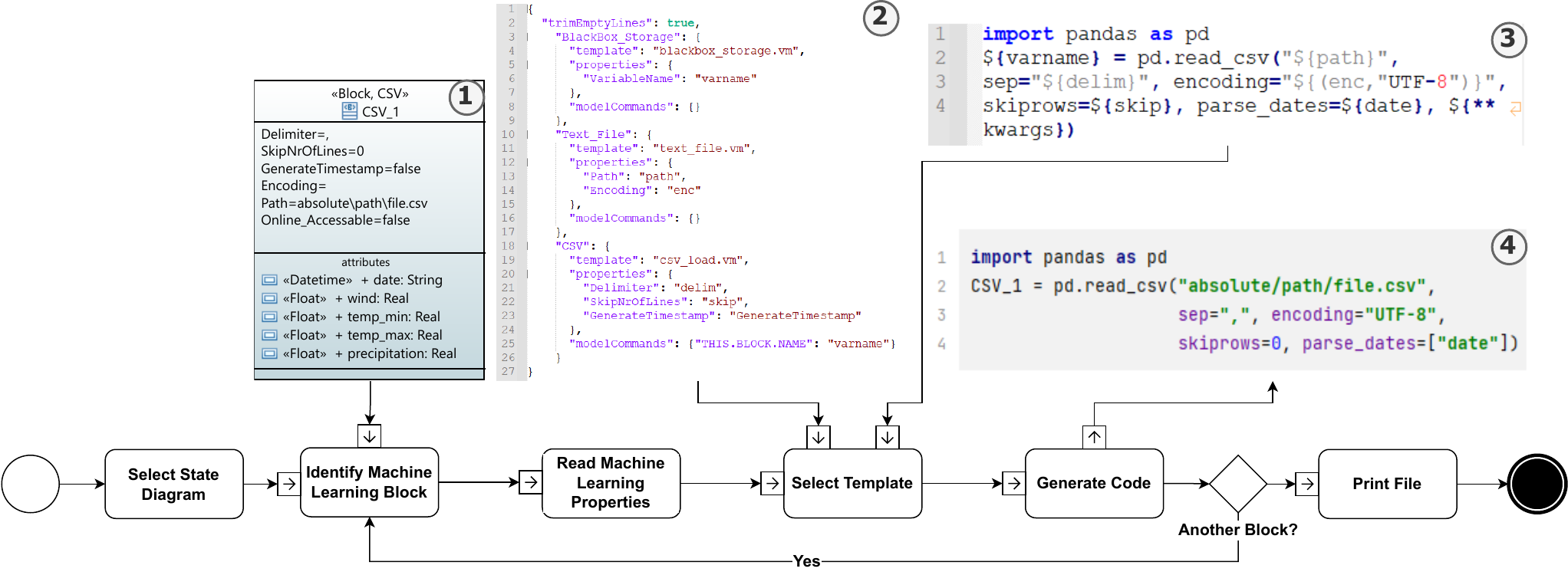}
    \caption{A sample model transformation to load a CSV File.}
    \label{fig:TransformationWorkflow}
\end{figure*}

To enable the decomposition of the defined SysML model, the here presented method relies on templates, defined as code snippets in a dedicated programming language, such as Python, and a mapping configuration that allows to identify a template based on a stereotype.
The purpose of the template-based approach is to enable extendability and maintainability without the necessity to make changes in the model transformation.
Additionally, an exchange of the template can be used to derive code within another programming language, such as JAVA or R.
Figure \ref{fig:TransformationWorkflow} depicts the generic method to generate machine learning code based on templates in a flow-diagram aligned workflow, aligned with a sample model transformation depicted as images on top of the figure.
The transformation applies the following subsequent steps:
\begin{enumerate}
    \item A state diagram is provided as input, referencing each machine learning subtask formalized using stereotypes and blocks
    \item For each of the states, which are provided in ascending order, the machine learning blocks are identified.
    \item Based on the unique stereotype name, a template is selected.
    \item Stereotype and block attributes \circled{\small1} are mapped to the template \circled{\small3} using a mapping configuration \circled{\small2} to generate a code snippet \circled{\small4} (see Figure \ref{fig:TransformationWorkflow}
    \item A file is generated representing the executable code snippets in the correct execution order. In the actual prototype implementation, a Jupyter Notebook is generated.
\end{enumerate}

The model transformation in Figure \ref{fig:TransformationWorkflow} is slightly simplified and omits the step of an intermediate transformation. 
In the following, this intermediate transformation is introduced in Section \ref{sec:intermediateModel}.
Next, the composition of the templates with placeholders is introduced.
Finally, the mapping configuration is introduced, focusing on model commands.

\subsection{Intermediate Model}\label{sec:intermediateModel}
The purpose of the intermediate model is to extract information from the SysML model and to merge the state diagram information with the linked blocks.
The source metamodel is a SysML model, and the target metamodel is a custom one, referred to as ``block context'' in the following.
The block context consists of the following parts: 

First, a reference to the original block in the SysML model to allow change tracking and to potentially enable synchronizing changes in the generated code with the original model.

Second, a list of rich-text blocks that can be rendered as text before a code block, modeled as so-called owned comments in the SysML model.
Note that rich-text annotations are represented as text block cells in the actual implementation. A dedicated format for Jupyter Notebooks and must be considered separately for other environments or programming languages.

Third, references to connected block contexts based on the qualified name, a unique identifier for named SysML elements.
Due to the uniqueness of the qualified name, it can be used as an identifier for attributes or blocks.

Fourth, a list of block and stereotype attributes with their values. 
If a value is a primitive type, the value is used. 
Otherwise, the qualified name is stored and translated to a value when assigned.

Finally, an integer represents the execution order in the state diagram.
The transformation is executed for each block connected to a state and for each block connected to such a block.
Care is taken to prevent the multiple execution of the transformation for the same block more than once by tracing the unique identifiers of a block.

\subsection{Code Snippet Template Definition}
The templates defining code snippets are defined in textual editors.
Particularly, a template consists of formatted plain-text with various placeholders filled with property values from the stereotypes during the code generation.
The marker \circled{\small3} in Figure \ref{fig:TransformationWorkflow} depicts a sample of a template with all possible types of variables, which are:

\begin{enumerate}
    \item Standard variables are highlighted with \textit{\$\{variable name\}}. In this case, the attribute is mandatory and has to be set in the model.
    \item An optional variable that alternatively is set with a default value, indicated with a default value in the variable definition \textit{\$\{(variable name, default value)\}}.
    \item Arbitrary other attributes.
\end{enumerate}

Since a function in a code snippet can have countless attributes, not all attributes can be defined in a stereotype, and it would not make sense due to the complexity for the user.
Therefore, additional properties can be added to a block instance without being defined in the stereotype.
These additional properties are added to a specific position in the template indicated by an anchor-indicator \textit{**kwargs}. 
For an additional property to be used for the template function, the name of the additional property must be similar to the parameter name of the corresponding programming language function, but with two asterisks after it, e.g., if a parameter of a chart printing function in Python calls \textit{X-Axis Name}, the attribute in the block must be named \textit{**X-Axis Name}.
The double-stared unforeseen attributes are rendered to the template in the following format $attribute\_name=attribute\_value$ without the double-stars.
If a template requires two \textit{**kwargs}, the transformation must be adapted, or two sub-stereotypes must be used.

\subsection{Mapping Configuration}\label{sec:mapconfiguration}
A mapping configuration in \circled{\small3} in Figure \ref{fig:TransformationWorkflow} illustrates the content of a mapping between a stereotype and a code snippet template using JSON file format.
The definition of JSON mapping is depicted in Listing \ref{lst:mappingSyntax}.

\begin{lstlisting}[caption={JSON Mapping Structure},label={lst:mappingSyntax}, language=json, float, numbers=left, stepnumber=1]
{
  "trimEmptyLines": <true||false>,
  "constants": {
    "<TemplateVariableName>": "<ConstantValue>",
    ...
  },
  "stereotypeMappings": {
    "<StereotypeName>": {
      "template": "<TemplateName>",
      "properties": {
        "<stereotypeAttributeName>": "<TemplateVariableName>",
        ...
      },
      "modelCommands": {
        "<ModelCommandKeywordCombination>": "<TemplateVariableName>",
        ...
      }
    },
  "nameMappings": {
    "<BlockName>": {
      "template": "<TemplateName>",
      "properties": {
        "<PropertyOrStereotypeAttributeName>": "<TemplateVariableName>",
        ...
      },
      "modelCommands": {
        "<ModelCommandKeywordCombination>": "<TemplateVariableName>",
        ...
      }
    }
  }
}
\end{lstlisting}

The mapping configuration is defined as follows:

First, the mapping allows defining whether empty lines shall be trimmed during the generation of the Jupyter Notebook (Line 2 in Listing \ref{lst:mappingSyntax}).

Second, the definition of constant values allows reusing specific strings as static text, e.g., as a global variable for all templates (Line 3-6 in Listing \ref{lst:mappingSyntax}).

The stereotype mapping (Line 7-18 in Listing \ref{lst:mappingSyntax}) allows specifying which template to use for a stereotype. The stereotype mapping (Line 10-13 in Listing \ref{lst:mappingSyntax}) defines the mapping of stereotype properties to template variables. 

A command can be defined (Line 14-17 in Listing \ref{lst:mappingSyntax}) and mapped to a variable by using the following keywords to collect information:

\begin{enumerate}
    \item \textbf{THIS}: the information can be found on the block with the stereotype
	\item \textbf{CONNECTED[Name="", Nr=0, StereotypeName="", AttributeValue={"AttributeName": ""}, OUTPUT\_Name=""]}: the information can be found on an associated block based on a search query, e.g. CONNECTED[Name="Sensor\_Log"] for \textit{Format\_Date} in Fig. \ref{fig:SampleApplicationML}
    \item \textbf{BLOCK}: the information is stored on the block directly
	\item \textbf{STEREOTYPE["StereotypeName"]}: the information is stored on a specifically applied stereotype (blocks can inherit from multiple stereotypes)
    \item \textbf{NAME}: the information is the name of the block specified by the preceding keywords
	\item \textbf{ATTRIBUTES}: the information is a list of attributes defined in a specific block
	\item \textbf{STEREOTYPEofATTRIBUTE["AttributeName"]}: the information is stored in a data stereotype of an attribute, e.g. \textit{Datetime} stereotype of the \textit{date} attribute of the \textit{Sensor\_Log} block in Fig. \ref{fig:SampleApplicationML}
    \item \textbf{OUTPUT}: the information is the last declared variable name of the template, which refers to the block specified by the preceding keywords
\end{enumerate}

The command's syntax consists of at least three keywords, separated by a period.
The first keyword is either \textit{THIS} or \textit{CONNECTED} with a selector to choose the correct connected block.
The second keyword is either \textit{BLOCK} if the information is directly stored on the block or \textit{STEREOTYPE} with a parameter specified for the stereotype name if it does not belong to the block itself. 
The third parameter is depicted in the enumeration list of keywords above with the item numbers 5-8. 
After the last keyword, it is always possible to select a value if the result is a list using square selector $[Nr.]$. 
After the \textit{ATTRIBUTES} and \textit{STEREOTYPEofATTRIBUTE}, optionally \textit{ATTRIBUTES} or \textit{STEREOTYPEofATTRIBUTE} can be defined again to dig deeper into specific information. 
The \textit{OUTPUT} value is one of the essential values to connect a code block with the result of a previous one.

If a specific mapping is only applied to a specific block, name mapping can be used (Line 19-31 in Listing \ref{lst:mappingSyntax}).
Name mapping is similar to stereotype mapping, but it specifies the input model block via the block name instead of the stereotype name.
The only difference is that properties can also be defined on the block without being defined on the stereotype.
Name mappings take precedence over stereotype mappings if both apply for a block.

\subsection{Composition of Code Snippets}\label{sec:Jupyter}
Based on the generated code snippets and the defined execution order of the snippets, an executable file can be generated.
The method presented in this work is implemented for Jupyter Notebook.
For this, the following steps for composition are conducted:

\begin{enumerate}
    \item Rich-text information modeled as owned or applied comment is directly converted to a Jupyter rich-text cell.
    \item The generated templates are put in a source-code cell. Each block context (intermediate model) from the state machine gets one source code cell and, optionally, one rich-text cell.
\end{enumerate}

The code snippets are analyzed for \textit{"from ... import ..."} or \textit{"import ..."} lines of code to increase the readability and reduce potential errors due to multiple inputs of modules required.
These lines are cut out and inserted in the first code cell on top of the Jupyter Notebook file.

After all block contexts are iterated over, the cells are put together as a single file, leading to an executable Jupyter Notebook file.
Finally, the syntax is validated, so the execution is ensured.
If the syntax is incorrect, the user is notified, but the task is still defined as completed.
The validation for semantics is considered out of scope.

\section{Evaluation}
The evaluation of the presented method aims to assess the feasibility and applicability of the method for generating executable machine learning code.

In the following, we present the case study used for the evaluation with the used artifacts from an open dataset.
Additionally, an excerpt of the generated artifact is presented. The comprehensive results and generation of code is available online\footnote{\url{https://github.com/sraedler/MDE_for_ML_Generation}}.

\subsection{Case Study and Artifacts}
As of\cite{brambilla_model-driven_2017}, two approaches can be followed to implement a model transformation, 1) using current high-level programming languages, APIs, and frameworks or 2) relying on MDE principles and dedicated languages such as ATL\footref{footnote} and Epsilon\footref{footnote}.
This evaluation uses traditional programming paradigms and the well-known high-level programming language JAVA.

The dataset for the evaluation is based on an open dataset\footnote{\url{https://www.kaggle.com/datasets/ananthr1/weather-prediction}} to predict weather forecasts based on sensor data from a weather station.
The scenario of a weather forecast based on weather station data is suitable for application in the engineering domain because the data comprises multiple sensors with different timestamps and sampling rates.
Additionally, the use of temperature or humidity sensors is also relevant in manufacturing specific components and the resulting quality.
The model transformation concept is wholly decoupled from data-driven engineering and could therefore be evaluated for any machine learning problems.

The modeling of the machine learning tasks is depicted in a previous publication\cite{radler_integration_2022}, extended with various comments to support the readability of the generated code.
Further details, such as a representation of the model as images or the transformation result of the evaluation, are shown in the artifacts available online\footnote{\url{https://github.com/sraedler/MDE_for_ML_Generation}}.

\subsection{Results}
This section depicts the results from the model transformation applied to the model in\cite{radler_integration_2022}.

\begin{figure}
    \centering
    \includegraphics[width=\columnwidth]{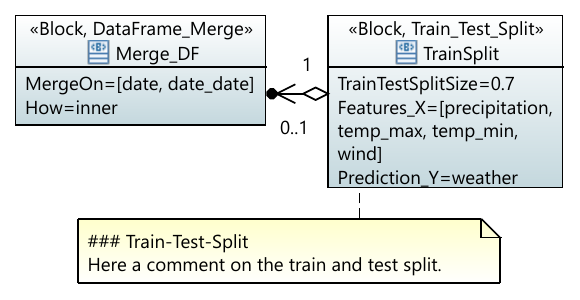}
    \caption{Sample input model.}
    \label{fig:train_model}
\end{figure}

\begin{figure}
    \centering
    \includegraphics[width=\columnwidth]{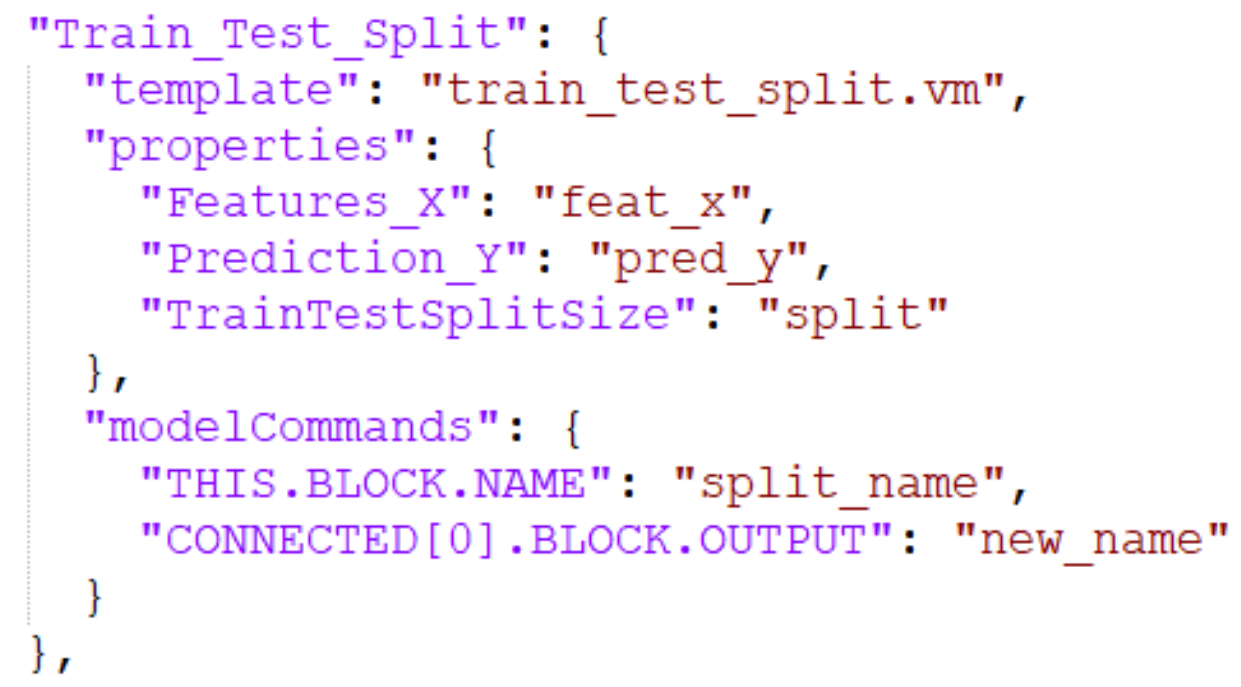}
    \caption{Mapping configuration.}
    \label{fig:train_mapping}
\end{figure}

\begin{figure}
    \centering
    \includegraphics[width=\columnwidth]{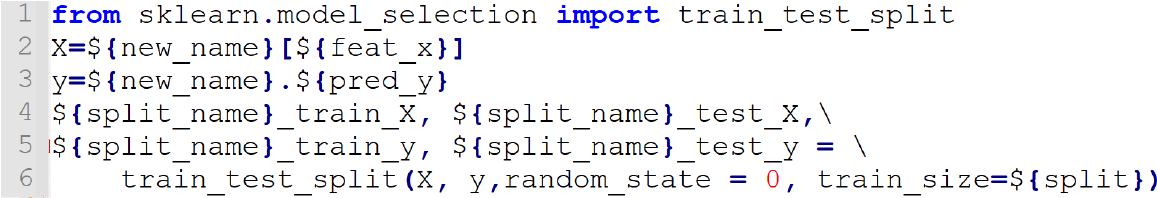}
    \caption{Template for the Train\_Test\_Split stereotype.}
    \label{fig:train_template}
\end{figure}

\begin{figure}
    \centering
    \includegraphics[width=\columnwidth]{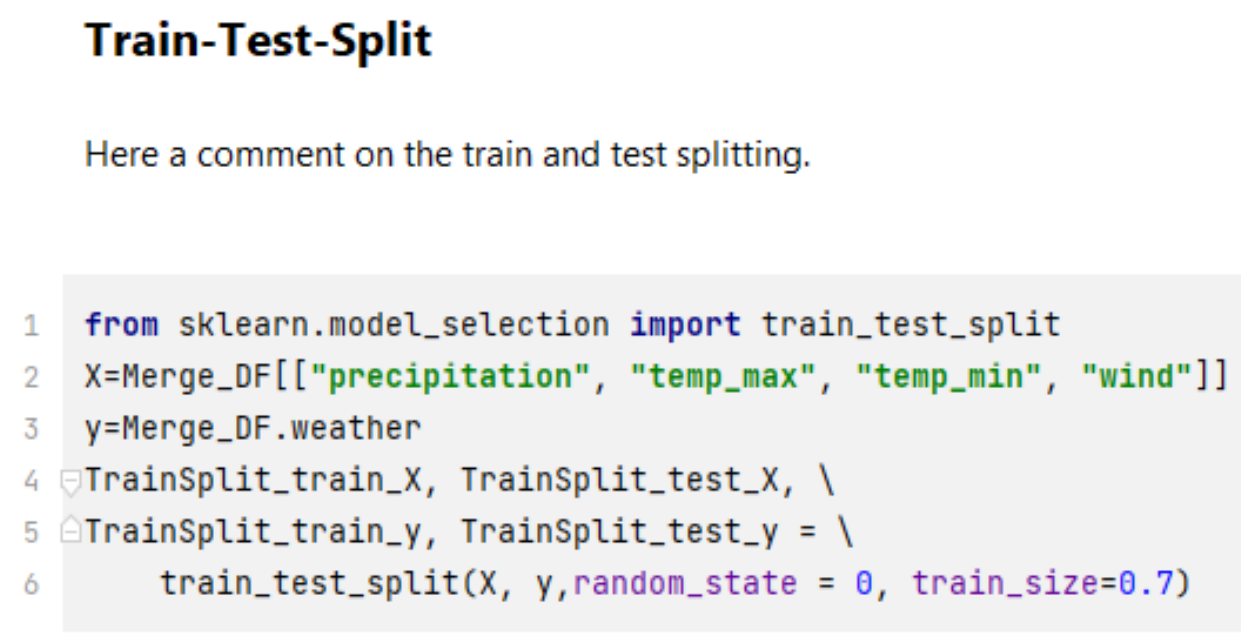}
    \caption{Result of the code generation.}
    \label{fig:train_result}
\end{figure}

Figure \ref{fig:train_model} to Figure \ref{fig:train_result} depict the four parts of the developed model transformation.

Figure \ref{fig:train_model} depicts two blocks with stereotype properties defined and a block comment connected to a block, which is further used in the final Jupyter Notebook as Rich-Text Cell.
The \textit{TrainSplit} block is defined only by stereotype attributes.
Additional attributes for hyper-parameter tuning, etc. are not defined.
The composition indicates that the \textit{Merge\_DF} block is an input value for the \textit{TrainSplit} function.
Therefore, it is accessible through the \textit{modelCommand} functionality defined in Listing \ref{lst:mappingSyntax}.

To enable the mapping from the input model in Figure \ref{fig:train_model} to the output in Figure \ref{fig:train_result}, a mapping configuration as defined in Figure \ref{fig:train_mapping} and a template as depicted in Figure \ref{fig:train_template} is required.
The mapping configuration assigns a stereotype \textit{Train\_Test\_Split} to a template with a name and, potentially, a path if sub-folders are used in the given structure.
Each stereotype property is defined within the template's properties, whereas the left side of the assignment is the original variable in the stereotype and the right side is the placeholder in the template.
The mapping defines two \textit{modelCommands}, i.e., the first to get the name of the actual block and the second one to collect the output variable of the first connected block.

Figure \ref{fig:train_template} illustrates a sample code snippet for a machine learning function, more precisely, a template for the train-test-split.
Within each template, necessary imports must be defined, and arranged at the end of the code generation, as defined in Section \ref{sec:Jupyter}.

Figure \ref{fig:train_result} depicts the generated code based on the template and the input model attributes.
As it can be seen, the formatting is aligned with the template in Figure \ref{fig:train_template}.

\section{Discussion}
This section discusses the introduced code generation method for machine learning based on model transformation and SysML.
First, general advantages and disadvantages are discussed.
Next, quality attributes of model transformation are discussed to allow an assessment of code generation.
Finally, potential future work is presented.

\subsection{Advantages and Disadvantages}
Using model transformation to decompose formalized machine learning tasks is beneficial in several ways.
First, it reduces the programming effort required for machine learning and consequently reduces the effort for rarely available data scientists\cite{radler_survey_2022}.
In addition, it allows the formalized knowledge in the model to be validated from an implementation point of view.
Validated knowledge enables the creation of a proven machine learning model library, leading to standardization of machine learning implementation within an organization's infrastructure.
This potentially favors the creation of machine learning tasks without profound programming knowledge.

Nevertheless, the method can be costly for small programs and too complex and cumbersome for large-scale problems.
One reason is the initial effort required to create and validate templates.
However, the resulting templates lead to standardization and can thus be reused in multiple projects, which becomes an advantage in future projects.
Another reason is that traceability can suffer with larger and more complex data preprocessing steps, requiring additional documentation.
However, these problems are more of the nature of formalization than model transformation.
Finally, it should be mentioned that the transformation is currently only directional, which does not allow changes in the generation code to be synchronized with the model.

\subsection{Quality Attributes of Model Transformation}
Quality attributes of model transformation can be distinguished in direct assessment, which is the actual assessment of the model transformation and its properties, and indirect by analyzing the input and output artifacts, e.g., metamodels\cite{van_amstel_right_2010,van_amstel_assessing_2012}.
Furthermore, a distinction is made between internal quality, which focuses on development and maintenance, and external quality, which focuses on compliance with requirements and performance\cite{van_amstel_right_2010,van_amstel_assessing_2012}.
In the following, direct internal quality attributes are discussed.
Although various metrics are available to assess these quality dimensions, a qualitative discussion is chosen.
The metrics are adapted to the transformation language used, which is not applicable here as the implementation is not based on a transformation language but on a traditional programming language\cite{van_amstel_metrics_2010}.

\subsubsection{Understandability}
The effort required to understand the purpose of the model transformation\cite{van_amstel_metrics_2008}.

The model transformation is easy to understand because a high-level programming language is used for the implementation, which can be adopted by most programmers. In contrast, the use of specific transformation languages such as ATL or Epsilon is less common and therefore the concept needs to be learned and understood.
Moreover, the overall concept of mapping model artifacts using a configuration in JSON file format is a simple technique with typical concepts known from programming.
A possible argument for weaknesses is that the programming of the JAVA code could be more complex to understand than MDE techniques for transformation.
However, these problems are more related to software engineering than to model transformation.

\subsubsection{Modifiability}
The effort required to adopt a model transformation to provide other or additional functions\cite{van_amstel_metrics_2008}.

The effort for modifications is potentially small because 1) the input metamodel can be adapted, and the concept of mapping attributes to a template is simple 2) the mapping configuration is highly customizable and can be adapted without deep programming experience, and 3) the output templates are small code fragments that can be formulated in any programming language.
In addition, any functions can be added from the programming perspective by adding additional templates or stereotypes.
The mapping already provides \textit{modelCommands}, allowing the collection of specific attributes or related information.
Even if more complex extensions are required, such as inserting security-related code to authenticate users, this can be adapted due to the use of the high-level language JAVA. 

\subsubsection{Reusability}
The extent to which parts of a model transformation can be reused by other (related) model transformations\cite{van_amstel_metrics_2008}.

Due to the possibility to exchange the output templates, the transformation can be applied to any textual programming language that enables machine learning and can be assembled from small code fragments.
Similarly, the concept of transformation can be used for any other model-to-code generation that can be broken down into small code fragments, as it is simply a mapping mechanism between input stereotype and output template.

\subsubsection{Modularity}
The extent in which a model transformation is systematically
separated and structured\cite{van_amstel_metrics_2008}.

Modularity is given in two aspects. 
First, stereotypes can be arbitrarily organized as long as they inherit from the core \textit{ML} stereotype.
Second, output templates can be stored in folders to structure templates.
However, the method does not allow defining a mapping only for a specific subset of functions.
Therefore, always a single JSON is required to represent the mapping configuration.
Nevertheless, extending the method to include the ability to parse multiple JSON files for mapping configuration is possible with little effort, allowing for complete separation and modularization of certain aspects of transformation.

\subsubsection{Completeness}
The extent to which a model transformation is fully developed in relation to the requirements\cite{van_amstel_metrics_2008}.

Completeness is conformance to requirements, which can be separated into functional or non-functional requirements of the model transformation.

The functional requirements for the model transformation can be summarized as the ability to generate executable machine learning code, which is given as of the first evaluation.

From a non-functional perspective, aspects such as generation performance must be evaluated.
Due to the early stage of development, performance assessment using Big-O-notation is not reliable, as extensions are potentially required that distort the estimate.
For this reason, the non-functional requirements are not yet assessed.

\subsubsection{Consistency}
The extent in which a model transformation is implemented
in a uniform manner\cite{van_amstel_metrics_2008}.

Because of the small programming code that compiles the input, mapping configuration, and templates, consistency is not a main quality criterion in this method, as it would be if ATL or Epsilon were used.
Therefore, this criterion is not further discussed.

\subsubsection{Conciseness}
The extent to which a model transformation is free of superfluous elements\cite{van_amstel_metrics_2008}.

Due to the high entanglement of the mapping configuration and the code templates, superfluous elements are barely available.
Additionally, the functionality to add arbitrary attributes to the generation using \textit{**kwargs} reduces the number of superfluous elements.
However, elements can be created during the modeling, or unnecessary templates can be defined.
However, these expressions are part of the nature of the application rather than a weakness of the model transformation.

\subsection{Future Work}
Future work involves implementing improvements and validating the method within user studies to prove its applicability in industrial projects.
Additionally, the systematic backflow of results from machine learning to the SysML model requires to be implemented to allow to use the yielded results in further model-based systems engineering methods.
Similarly, it is beneficial if changes in the Jupyter Notebook can be traced back to the model so that synchronization and an authoritative source of truth\footnote{\url{https://www.omgwiki.org/MBSE/doku.php?id=mbse:authoritative_source_of_truth}} can be achieved.
With respect to this, the actual transformation traces the model elements, allowing the identification of the origin, and within the Jupyter Notebook, unique block markers can be used to map the changes to the model elements.
However, profound changes require a mechanism to generate further blocks or adapt templates.

\section{Conclusion}
This work presented a model transformation to facilitate machine learning applications using model-based techniques based on the general-purpose modeling language SysML.
The goal of the code generation is to enable standardization of machine learning code within a company and allow to reuse formalized knowledge on machine learning tasks.
The code generation is enabled by generic templates providing concise code snippets that are mapped using a mapping configuration defined in JSON file format to stereotypes or specific blocks in the SysML formalization.
The generated executable code enables the validation of the formalized machine learning tasks in the SysML model.
The method is validated in a case study and artifacts are made available online.
Through the study results, RQ1: "What model characteristics can be used to derive machine learning models to enable model-driven engineering automatically?" can be answered by using stereotypes to identify features. Based on this, templates made specifically on stereotypes are selected, and attributes of the stereotypes are inserted into the template.
Consistent with the answer to RQ1, RQ2: "What means of software engineering allows to extend and maintain the machine learning code derivation without profound model transformation changes?" can be answered by choosing a code generation that builds on a mapping configuration with templates and stereotyped definition.

By extending stereotypes and adapting or adding new templates, the method can be extended without changes in code generation.
Moreover, this allows the target programming language to be selected based on the defined templates without changing the model transformation.

Future work will include elaborating an information backflow from the derived and potentially changed code to the SysML model to introduce a single source of truth. 
Further, the method will be validated in a user study to improve its applicability in practice and streamline the research agenda of model-driven engineering methods for machine learning.

\bibliographystyle{IEEEtran}
\bibliography{references}
\end{document}